\documentclass[useAMS,usegraphicx,usenatbib]{mn2e}

\usepackage{graphicx}
\usepackage{times}
\usepackage{amssymb}
\usepackage{amsmath}
\newif\ifAMStwofonts
\AMStwofontstrue

\def\chandra{{\it Chandra}}
\def\hst{{\it HST}}
\def\hubble{{\it Hubble}}
\def\hubblefull{{\it Hubble Space Telescope}}
\def\xmm{{\it XMM-Newton}}

\def\spitzerfull{{\it Spitzer Space Telescope}}
\def\spitzer{{\it Spitzer}}

\def\ms{{\rm\thinspace Ms}}

\def\a{{\rm\thinspace\AA}}

\def\um{{\hbox{$\rm\thinspace \umu m$}}}

\def\amsq{\hbox{$\rm\thinspace arcmin^{2}$}}
\def\degsq{\hbox{$\rm\thinspace deg^{2}$}}

\def\pdegsq{\hbox{$\rm\thinspace deg^{-2}$}}

\def\pcmsq{\hbox{$\rm\thinspace cm^{-2}$}}

\def\kev{{\rm\thinspace keV}}

\def\ergpcmsqps{\hbox{$\rm\thinspace erg~cm^{-2}~s^{-1}$}}

\def\ergps{\hbox{$\rm\thinspace erg~s^{-1}$}}

\def\kevpcmsqpspsrpkev{\hbox{$\rm\thinspace keV~cm^{-2}~s^{-1}~sr^{-1}~keV^{-1}$}}

\def\snr{\hbox{\rm SNR}}

\title[] {Can the unresolved X-ray background be explained by emission from the
  optically-detected faint galaxies of the GOODS project?}  \author[M. A.
Worsley et al.]  {\parbox[]{6.in}
  {M.~A. Worsley,$^{1}$ A.~C. Fabian,$^{1}$\thanks{E-mail: acf@ast.cam.ac.uk} F.~E.~Bauer,$^{2}$ D.~M.~Alexander,$^{1}$ \\ W.~N.~Brandt$^{3}$ and B.~D.~Lehmer$^{3}$}\\\\
  \footnotesize
  $^{1}$Institute of Astronomy, Madingley Road, Cambridge CB3 0HA\\
  $^{2}$Department of Astronomy, Columbia University, 538 W 120th Street, New York, NY 10027, USA\\
  $^{3}$Department of Astronomy and Astrophysics, 525 Davey Laboratory, Pennsylvania State University, University Park, PA 16802, USA\\
}

\begin{document}
\maketitle

\label{firstpage}

\begin{abstract} 
  The emission from individual X-ray sources in the \chandra\ Deep Fields and
  \xmm\ Lockman Hole shows that almost half of the hard X-ray background above
  \hbox{$6\kev$} is unresolved and implies the existence of a missing population
  of heavily obscured active galactic nuclei (AGN). We have stacked the
  \hbox{$0.5$--$8\kev$} X-ray emission from optical sources in the Great
  Observatories Origins Deep Survey (GOODS; which covers the \chandra\ Deep
  Fields) to determine whether these galaxies, which are individually undetected
  in X-rays, are hosting the hypothesised missing AGN. In the
  \hbox{$0.5$--$6\kev$} energy range the stacked-source emission corresponds to
  the remaining \hbox{$10$--$20$} per cent of the total background -- the
  fraction that has not been resolved by \chandra. The spectrum of the stacked
  emission is consistent with starburst activity or weak AGN emission. In the
  \hbox{$6$--$8\kev$} band, we find that upper limits to the stacked X-ray
  intensity from the GOODS galaxies are consistent with the \hbox{$\sim40$} per
  cent of the total background that remains unresolved, but further selection
  refinement is required to identify the X-ray sources and confirm their
  contribution.
\end{abstract}


\section{Introduction}
\label{intro}

Most of the extragalactic X-ray background (XRB) has been conclusively shown to
be the integrated emission from discrete sources, in particular, the accretion
light from Active Galactic Nuclei (AGN). The \hbox{$1$--$10\kev$} XRB has a
spectral slope of \hbox{$\Gamma=1.4$} with a \hbox{$5$--$10$} per cent spread in
measurements of the normalisation \citep[see e.g.][]{revnivtsev03,deluca04}, a
large amount of which is due to field-to-field variations \citep{barcons00}.
Emission from sources resolved by \chandra\ in the broad \hbox{$0.5$--$2\kev$}
and \hbox{$2$--$10\kev$} bands is able to account for \hbox{$80$--$90$} per cent
of the XRB \citep{mushotzky00,giacconi02,alexander03a,bauer04}, leading to some
claims that the origin of the background has been solved. Recent analysis of
very hard (\hbox{$>5\kev$}) X-ray data, however, as well as growing evidence
from infrared and submillimetre studies, indicate that a substantial number of
hard X-ray emitting AGN may remain to be found.

Although the resolved XRB in the broad \hbox{$2$--$10\kev$} range is high, this
does not imply that the background is accounted for at \hbox{$10\kev$}, since
the fraction is dominated by counts in the \hbox{$2$--$6\kev$} range to which
telescopes such as \chandra\ are most sensitive \citep{worsley05}. Narrow energy
band source-stacking in the \xmm\ Lockman Hole (XMM-LH) and \chandra\ Deep
Fields North (\hbox{CDF-N}) and South (\hbox{CDF-S}) indicate that the resolved
fraction of the XRB decreases from \hbox{$\sim80$--$90$} per cent over
\hbox{$2$--$8\kev$}, to \hbox{$\sim60$} per cent over \hbox{$6$--$8\kev$}, and
only \hbox{$\sim50$} per cent \hbox{$>8\kev$} \citep{worsley04,worsley05}. These
analyses derived the resolved fraction by summing, in narrow energy bands, the
X-ray fluxes from all individually-detected sources. The same source list was
used for each narrow band considered, regardless of whether or not a source is
explicitly detected in that band (most, but not all, sources were actually
detected in the soft \hbox{$\lesssim2\kev$} range; \citealp[refer to][and
references therein]{worsley04,worsley05}). The unresolved background component
has a spectral shape that is consistent with an unresolved population of
highly-obscured AGN at redshifts \hbox{$\sim0.5$--$1.5$} and with absorption
column densities of
\hbox{$\sim10^{23}$\thinspace--\thinspace$5\times10^{24}\pcmsq$}. Since the
sources have not been individually detected in X-rays, their intrinsic,
de-absorbed luminosities are probably \hbox{$\lesssim5\times10^{43}\ergps$}
(rest-frame \hbox{$2$--$10\kev$}), though this depends on important assumptions
about their space density and absorption column densities.

The steeply-rising X-ray source number counts in the \hbox{$5$--$10\kev$} band
show no evidence for flattening towards fainter fluxes
\citep{hasinger01,rosati02,baldi02}. Since the average spectral slope of sources
is known to be a strong function of flux -- becoming progressively harder with
decreasing flux \citep{ueda99,fiore03,alexander03a,streblyanska03} -- this also
suggests a large, undiscovered population of faint, hard X-ray sources.

Locally, highly obscured AGN outnumber their unobscured counterparts by a factor
of \hbox{$\sim4$:$1$}. The three nearest luminous AGN -- Cen A, NGC4945 and
Circinus -- all have column densities \hbox{$>10^{23}\pcmsq$}, with NCG4945 and
the Circinus galaxy being Compton thick \citep{matt00}. Population synthesis
models \citep[e.g.][]{gilli01,franceschini02,gandhi03,ueda03,comastri04} require
large numbers of highly obscured AGN in order to account for the \hbox{$30\kev$}
peak in the XRB spectrum \citep{setti89}. If the ratio of obscured to unobscured
AGN remains high (e.g. \citealt{gilli01} predict \hbox{$10$:$1$}) out to high
redshift, then a large number of faint, obscured sources could remain undetected
in deep X-ray observations. At redshifts \hbox{$z\gtrsim1$} recent, careful
studies suggest the obscured:unobscured ratio is indeed high (e.g.
\citealt{alexander05} find \hbox{$\sim6$:$1$}).

Highly obscured AGN can show little or no soft X-ray emission due to
photoelectric absorption, particularly if the covering fraction of the obscuring
material is high and the fraction of reflected light is low. Current X-ray
telescopes have been able to detect many low-redshift obscured AGN, as well as
examples of type II quasars \citep[e.g.][]{norman02,stern02,gandhi04}, yet
sensitivity to faint, hard sources is still limited. Highly obscured sources
with Seyfert luminosities (e.g. the Circinus galaxy) are undetectable even in
the CDFs beyond redshifts of \hbox{$0.2$--$1$} \citep[see Fig.~6 in][]{brandt05}
-- the distance at which the majority of unobscured AGN are found. Submillimetre
and infrared observations \citep[see
e.g.][]{alexander03b,alexander05,treister04}, may be the key to identifying the
highly absorbed sources through the reprocessed AGN emission emerging at these
wavelengths. IR colour selection using \spitzer\ data should be able to
disentangle starburst emission from the re-radiated AGN emission
\citep{alonso-herrero04,lacy04,lacy05}.

The Great Observatories Origins Deep Survey (GOODS) \citep{dickinson03} is an
ongoing campaign of multiwavelength deep-field observations comprised of
existing and planned surveys. The programme is being carried out in two fields,
each \hbox{$\sim10\rm{'}\times16\rm{'}$} in size. The GOODS-N field is contained
within the \hbox{$\sim2\ms$} \hbox{CDF-N} \citep{alexander03a} and includes the
\hubble\ Deep Field North \citep{williams96,ferguson00}. The GOODS-S field lies
inside the \hbox{$\sim1\ms$} \hbox{CDF-S} \citep{giacconi02} and contains the
\hubblefull\ (\hst) Ultra Deep Field observation. As part of the GOODS project
the \spitzerfull\ has also carried out deep infrared surveys in the region from
\hbox{$3.6$--$24\um$} (PI M.~Dickinson) and the \hst\ Advanced Camera for
Surveys (ACS) has provided deep imaging of the fields in four broad wavebands
\citep{giavalisco04}, not to mention a host of observations from ground-based
telescopes.

We have performed X-ray stacking of optically-detected GOODS galaxies that have
not been individually detected in X-rays: we specifically ignore all the X-ray
sources that have already been individually detected in the CDFs -- stacking
analyses of these objects are covered in our earlier work
\citep[see][]{worsley05}. Since the GOODS survey probes optical galaxies to low
luminosity and high redshift, once the already-detected X-ray sources are
removed, the remaining `normal' galaxies are likely candidates to host the
hypothesised missing population of highly obscured AGN.

\section{Analysis}

\subsection{X-ray data and optically-detected sources}

The \hbox{$1.95\ms$} \hbox{CDF-N} covers a total of \hbox{$447.8\amsq$} and is
larger than the GOODS-N optical survey region, which is focused on the
\hbox{CDF-N} aim-point. \hbox{$3\sigma$} X-ray source detection sensitivities
are \hbox{$\sim2.5\times10^{-17}$} and \hbox{$\sim1.4\times10^{-16}\ergpcmsqps$}
in the \hbox{$0.5$--$2$} and \hbox{$2$--$8\kev$} bands respectively. We use the
main point-source catalogue \citep[see][]{alexander03a} which contains 503
sources. The \hbox{$0.94\ms$} \hbox{CDF-S} is also larger than the GOODS-S
optical region with an area of \hbox{$391.3\amsq$} and detection sensitivities
of \hbox{$\sim5.2\times10^{-17}$} and \hbox{$\sim2.8\times10^{-16}\ergpcmsqps$}
in the \hbox{$0.5$--$2$} and \hbox{$2$--$8\kev$} bands respectively. The main
point-source catalogue contains 326 sources \citep{alexander03a}.

The deep optical imaging of the GOODS regions, using the \hst\ Advanced Camera
for Surveys (ACS) is now available in the four non-overlapping photometric
filters: F435W ($B_{435}$), F606W ($V_{606}$), F775W ($i_{775}$) and F850LP
($z_{850}$) \citep{giavalisco04}. The \hbox{$5\sigma$} limiting magnitudes in
these four bands are 27.9, 28.2, 27.5 and 27.4 respectively (assuming an
aperture of \hbox{$0.5\rm{''}$}). We use the r1.1z version of the GOODS ACS
catalogue\footnote{Available at http://www.stsci.edu/science/goods/} which
contains 32,048 and 29,661 optically-detected sources in the entire GOODS-N and
GOODS-S regions respectively.

\subsection{X-ray image stacking}
\label{xrayimagestacking}

Our stacking procedure is similar to that taken by \citet{lehmer05} \citep[also
see][]{brandt01,nandra02}. At each optical source position we extract photon
counts from the X-ray images and exposure times from the X-ray exposure maps. We
used circular extraction apertures where a fixed 3-pixel radius
(\hbox{$1.476\rm{''}$}) was found to give the best signal-to-noise ratio (SNR)
compared to 2-pixel and 4-pixel apertures. To avoid contamination by X-ray
detected sources we did not stack any optical sources lying within three times
the 90 per cent encircled-energy fraction radius of an X-ray source
\citep{alexander03a}. We performed our stacking procedure in several narrow
energy bands: \hbox{$0.5$--$1$}, \hbox{$1$--$2$}, \hbox{$2$--$4$},
\hbox{$4$--$6$} and \hbox{$6$--$8\kev$}.

The background was estimated using a Monte Carlo approach. Each stacking
position was randomly shifted in RA and dec. up to \hbox{$60\rm{''}$} away from
its original position but was not allowed to overlap with the stacking aperture
of an optical source position, nor lie within three times the 90 per cent
encircled-energy fraction radius of an X-ray detected source. The Monte Carlo
procedure was carried out 10,000 times (except during the stacking of all the
sources in the catalogue, where only 1000 trials were computationally feasible
due to the large number of sources; see Table~\ref{stackingsummary}), and the
mean and variance of the background level determined.

We quantify the significance of the detection of X-ray emission from the stacked
objects in terms of the signal-to-noise ratio (\snr). This is given by
\begin{equation}
\snr = \frac{S-B}{\sqrt{B}},
\label{eqn:snr}
\end{equation}
where $S$ is the total number of stacked-source counts (i.e sources and
background) and $B$ is the total number of Monte Carlo stacked-background counts
(i.e. background only). For ease of reference, Table~\ref{probabilities} gives
the probability of a false detection (i.e. the probability of recording the
excess counts above background by chance alone), for various $\snr$s.

\begin{table}
\centering
\caption{The probability, $P_{\rm{false}}(\sigma)$, of falsely recording a given $\snr$, $\sigma$, in the stacking signal due to background fluctuations alone, when no sources are actually present. The probability of a genuine detection is thus \hbox{$1-P_{\rm{false}}(\sigma)$}. The values correspond to the one-tailed integrated Gaussian probabilities i.e. \hbox{$P_{\rm{false}}(\sigma)=(2\pi)^{-1/2}\int_{\sigma}^{\infty}\exp(-x^{2}/2)\thinspace \mathrm{d}x$}.}
\label{probabilities}
\begin{tabular}{cc}
\hline
\snr\ ($\sigma$) & $P_{\rm{false}}$ \\
\hline
$0.5$            & $0.309$             \\
$1$              & $0.159$             \\
$1.5$            & $0.0668$            \\
$2$              & $0.0228$            \\
$3$              & $1.35\times10^{-3}$ \\
$4$              & $3.17\times10^{-5}$ \\
$5$              & $2.87\times10^{-7}$ \\
\hline
\end{tabular}
\end{table}

\begin{table*}
\centering
\caption{Counts-to-flux conversion factors for the \hbox{CDF-N} and \hbox{CDF-S} data. The factors convert source count-rate (\hbox{$\rm{count~s^{-1}}$}) to flux (\hbox{$\rm{erg~cm^{-2}~s^{-1}}$}) and were computed assuming a \hbox{$\Gamma=1.4$} power-law spectrum inclusive of Galactic absorption. The factors include the Galactic absorption correction and also take into account the effects of absorption due to the \chandra\ molecular contamination. (See \citealt{alexander03a} for details.) The encircled energy fractions give the average fraction of source counts lying within the stacking aperture in each of the energy bands for the \hbox{CDF-N} and \hbox{CDF-S}. The values given were calculated using for the `all sources' cases with a 3-pixel (\hbox{$1.476\rm{''}$}) radius aperture.}
\label{ecfs_eefs}
\begin{tabular}{ccccc}
\hline
Energy band & \multicolumn{2}{c}{Counts-to-flux conversion factor}         & \multicolumn{2}{c}{Encircled energy fraction} \\
($\rm{keV}$)& \multicolumn{2}{c}{($10^{-11}~\rm{erg~cm^{-2}~count^{-1}}$)} & \multicolumn{2}{c}{(3-pixel radius aperture)} \\
            & \hbox{CDF-N} & \hbox{CDF-S} & \hbox{CDF-N} & \hbox{CDF-S} \\
\hline
$0.5$--$1$  & 0.69186      & 0.65327      & 0.705        & 0.704        \\
$1$--$2$    & 0.44369      & 0.46358      & 0.690        & 0.688        \\
$2$--$4$    & 1.4583       & 1.5830       & 0.653        & 0.652        \\
$4$--$6$    & 2.4763       & 2.7689       & 0.612        & 0.612        \\
$6$--$8$    & 8.9967       & 10.254       & 0.567        & 0.567        \\
\hline  
\end{tabular}
\end{table*}

Total count-rates were derived by dividing the total number of stacked counts by
the total stacked exposure time. These were then converted to fluxes using
counts-to-flux conversion factors from \citet{alexander03a}. We assumed a
\hbox{$\Gamma=1.4$} power-law plus Galactic absorption of
\hbox{$N_{\rm{H}}=1.3\times10^{20}$} and \hbox{$8.8\times10^{19}\pcmsq$} for the
\hbox{CDF-N} \citep{lockman03} and \hbox{CDF-S} \citep{stark92} respectively.
The conversion factors are given in Table~\ref{ecfs_eefs} and include the
necessary corrections for Galactic absorption and the low-energy absorption seen
in \chandra\ due to molecular contamination of the telescope's optical blocking
filters.

The size of the \chandra\ point spread function (PSF) varies strongly with
off-axis angle. The radius enclosing 50 per cent of the counts from a
point-source is only \hbox{$\lesssim0.5\rm{''}$} at the aim-point rising to
\hbox{$\gtrsim4\rm{''}$} at an off-axis angle of \hbox{$10\rm{'}$} (at
\hbox{$1.5\kev$}). Since the sky density of the optical sources is high enough
to result in overlap of the PSFs towards the edges of the fields we restricted
our analysis to the central \hbox{$4.5\rm{'}$} of the X-ray field to avoid
confusion between sources, and between sources and background. Stacking sources
in these central regions also enabled the use of a 3-pixel fixed aperture size,
which encloses the core of the PSF. The PSF remains fairly compact, and more
importantly, well-measured over these central regions and so the stacked counts
can be easily corrected for PSF effects. We used the data\footnote{{\it{Enclosed
      Count Fractions (ECF) using circular apertures based upon the SAOsac model
      of the Chandra PSF}} (2005 June 24) which is available at
  http://cxc.harvard.edu/cal/Hrma/psf/ECF/hrmaD1996-12-20hrci\_ecf\_N0002.fits}
from a circular parameterisation of the PSF in order to calculate and
correct-for the encircled-energy fraction at each aperture position and energy
band. The average encircled energy fractions in each of the energy bands for the
\hbox{CDF-N} and \hbox{CDF-S} are given in Table~\ref{ecfs_eefs}.

The total stacked-source fluxes were finally converted to intensity over the
solid angle on the sky in which the sources were stacked; in this case this is
simply the circular central \hbox{$4.5\rm{'}$} region (\hbox{$\sim63.6\amsq$}),
minus the area of the regions around X-ray detected sources which we excluded
from our analysis. The excluded solid angles are \hbox{$6.6$} and
\hbox{$4.3\amsq$} in the \hbox{CDF-N} and \hbox{CDF-S} respectively.

\subsection{X-ray Background Model}
\label{xrbmodel}

In order to calculate the fraction of the XRB which can be attributed to the
stacked sources we assume the XRB spectrum from \citet{worsley05}. This is a
\hbox{$1$--$8\kev$} power-law of photon index \hbox{$\Gamma=1.41$} and with a
\hbox{$1\kev$} normalisation of \hbox{$11.6\kevpcmsqpspsrpkev$} as observed by
\citet{deluca04}. Above \hbox{$8\kev$} the analytical model of \citet{gruber99}
is used, although renormalised to smoothly intercept the \hbox{$1$--$8\kev$}
power-law. Below \hbox{$1\kev$} we use a steeper power-law
(\hbox{$\Gamma\sim1.6$}), which intercepts the low-energy extragalactic XRB
measurements from \citet{roberts01}.

Whilst the slope of the XRB is well known in the \hbox{$1$--$8\kev$} range, the
normalisation measured by different instruments shows \hbox{$5$--$10$} per cent
variations. These measurements \citep[see
e.g.][]{vecchi99,lumb02,revnivtsev03,deluca04} are typically obtained over large
regions, typically several square degrees. The variations in XRB normalisation
can be explained in terms of uncertainties in the cross-calibration between
instruments, and the field-to-field variations arising from the discrete nature
of the sources making up the background. In pencil-beam fields such as the CDFs,
which only sample \hbox{$\lesssim0.1\degsq$}, field-to-field variations are at
least \hbox{$10$} per cent \citep{barcons00}, and further variations could be
due to true `cosmic variance' -- real differences in XRB level beyond those
simply arising due to sampling statistics. For example, spatial clustering on
the sky \citep[see e.g.][]{gilli03,gilli05} could be responsible for additional
variations in normalisation, whilst clustering in redshift space could create
additional variations in XRB spectral shape. Error in our XRB spectrum is taken
from the measurement errors quoted for the difference sources: we have not added
any additional uncertainty to allow for field-to-field variations.

\section{Results \& discussion}

\subsection{Optical source stacking}

Table~\ref{stackingsummary} summarises our stacking analyses of the
optically-detected GOODS sources. The \hbox{CDF-N} and \hbox{CDF-S} fields were
stacked separately to provide robustness. Stacking of all the optical sources in
the fields revealed highly significant (\hbox{$\sim10$--$30\sigma$}) detections
in the soft \hbox{$0.5$--$1$} and \hbox{$1$--$2\kev$} bands. Strong results are
seen in the \hbox{$2$--$4\kev$} band (\hbox{$7.3\sigma$} and \hbox{$7.8\sigma$}
for \hbox{CDF-N} and \hbox{CDF-S} respectively) and the broad
\hbox{$2$--$8\kev$} range (\hbox{$4.5\sigma$} and \hbox{$6.5\sigma$}). Detection
of flux in the very hard \hbox{$4$--$6\kev$} band is tentative at
\hbox{$2.6\sigma$} and \hbox{$2.0\sigma$}, although the combined \hbox{CDF-N}
and \hbox{CDF-S} detection is significant at the \hbox{$\sim99$} per cent level.
No significant detection is found in the \hbox{$6$--$8\kev$} band with only a
\hbox{$1.1\sigma$} signal in the \hbox{CDF-S} and a non-detection (i.e.
\hbox{$\snr <0$}) in the \hbox{CDF-N}. We would also note that we see no obvious
evidence for any spatial extension of the X-ray emission in the X-ray images
associated with the stacking analysis, although a more complete investigation
would be needed to pursue this further.

\begin{table*}
\centering
\caption{The results of stacking the emission from optically-detected, but individually X-ray undetected, GOODS sources. For the \hbox{CDF-N} and \hbox{CDF-S} analyses the results are shown for each of the different energy bands and also for the different $z_{850}$-band magnitude selections applied to the sources included in the stacking. If an excess of X-ray counts was detected then the detection \snr\ is quoted; `--' indicates non-detection, i.e. \hbox{$\snr <0$}.}
\label{stackingsummary}
\begin{tabular}{ccccccccc}
\hline
& Energy band        & \multicolumn{7}{c}{\snr\ ($\sigma$)} \\
& ($\rm{keV}$)       & \\
\hline
\multicolumn{1}{r}{\hbox{CDF-N} Sources:}     && $z_{850}<21$ & $z_{850}<22$ & $z_{850}<23$ & $z_{850}<24$ & $z_{850}<25$ & $z_{850}<26$ & all sources \\
\multicolumn{1}{r}{Number stacked:  }  && $117$  & $278$  & $571$  & $1160$ & $2132$ & $3720$ & $10,052$    \\
\hline
& $0.5$--$1$                           & $10.1$ & $13.8$ & $15.9$ & $17.1$ & $17.1$ & $16.2$ & $14.6$       \\
& $1$--$2$                             & $10.8$ & $20.0$ & $26.4$ & $28.2$ & $31.1$ & $29.8$ & $27.6$       \\
& $2$--$4$                             & $ 6.8$ & $ 8.2$ & $ 9.7$ & $ 8.4$ & $ 7.7$ & $ 8.3$ & $ 7.3$       \\
& $4$--$6$                             & $ 2.3$ & $ 2.9$ & $ 4.3$ & $ 3.2$ & $ 3.1$ & $ 3.2$ & $ 2.6$       \\
& $6$--$8$                             & --     & --     & $ 1.2$ & $ 1.0$ & $ 0.4$ & $ 0.2$ & --           \\
\hline
\multicolumn{1}{r}{\hbox{CDF-S} Sources:}     && $z_{850}<21$ & $z_{850}<22$ & $z_{850}<23$ & $z_{850}<24$ & $z_{850}<25$ & $z_{850}<26$ & all sources \\
\multicolumn{1}{r}{Number stacked:}    && $128$  & $289$  & $537$  & $1034$ & $1913$ & $3426$ & $9599$      \\
\hline
& $0.5$--$1$                           & $ 9.4$ & $ 9.9$ & $10.6$ & $11.5$ & $11.7$ & $12.2$ & $10.1$       \\
& $1$--$2$                             & $13.7$ & $17.8$ & $20.3$ & $21.9$ & $23.4$ & $23.3$ & $22.2$       \\
& $2$--$4$                             & $ 3.8$ & $ 5.4$ & $ 5.6$ & $ 7.7$ & $ 8.7$ & $ 6.7$ & $ 7.8$       \\
& $4$--$6$                             & $ 1.7$ & $ 1.4$ & $ 2.1$ & $ 1.7$ & $ 2.8$ & $ 1.5$ & $ 2.0$       \\
& $6$--$8$                             & $ 1.5$ & $ 1.9$ & $ 0.7$ & --     & $ 0.5$ & $ 0.7$ & $ 1.1$       \\
\hline
\end{tabular}
\end{table*}

\begin{table*}
\centering
\caption{The percentage of the total XRB flux in the signal from stacking the optically-detected, but individually X-ray undetected, GOODS sources. For the \hbox{CDF-N} and \hbox{CDF-S} analyses the results are shown for each of the different energy bands and also for the different $z_{850}$-band magnitude selections applied to the sources included in the stacking. In the case of non-detection (i.e. \hbox{$\snr <0$}), a $3\sigma$ upper limit is given. The `already resolved' column indicates the fraction of the XRB which has already been accounted for by X-ray detected sources \citep[see][]{worsley05}.}
\label{countssummary}
\begin{tabular}{cccccccccc}
\hline
& Energy band        & \multicolumn{8}{c}{Percentage of XRB flux}\\
& ($\rm{keV}$)       & & \\
\hline
\multicolumn{1}{r}{\hbox{CDF-N} Sources:} && $z_{850}<21$  & $z_{850}<22$  & $z_{850}<23$  & $z_{850}<24$  & $z_{850}<25$  & $z_{850}<26$  & all sources   & already   \\
\multicolumn{1}{r}{Number stacked:}&& $117$         & $278$         & $571$         & $1160$        & $2132$        & $3720$        & $10,052$      & resolved  \\
\hline
& $0.5$--$1$                        & $1.4 \pm0.3 $ & $2.8 \pm0.4 $ & $4.6 \pm0.7 $ & $7.1 \pm1.0 $ & $9.6 \pm1.3 $ & $12.1\pm1.6 $ & $17.7\pm2.5 $ & $86\pm11$ \\
& $1$--$2$                          & $0.8 \pm0.1 $ & $2.1 \pm0.2 $ & $4.0 \pm0.3 $ & $6.1 \pm0.4 $ & $9.1 \pm0.5 $ & $11.5\pm0.7 $ & $17.3\pm1.0 $ & $83\pm5 $ \\
& $2$--$4$                          & $1.3 \pm0.3 $ & $2.3 \pm0.5 $ & $3.9 \pm0.7 $ & $5.0 \pm0.9 $ & $6.2 \pm1.2 $ & $8.8 \pm1.6 $ & $12.6\pm2.6 $ & $91\pm8 $ \\
& $4$--$6$                          & $0.9 \pm0.6 $ & $1.7 \pm0.9 $ & $3.5 \pm1.2 $ & $3.8 \pm1.8 $ & $5.0 \pm2.3 $ & $6.8 \pm3.1 $ & $9.2 \pm5.1 $ & $92\pm13$ \\
& $6$--$8$                          & $<8.9$        & $<13.2$       & $5.2 \pm6.3 $ & $6.4 \pm9.2 $ & $3.2 \pm12.4$ & $1.9 \pm16.4$ & $<81.9$       & $56\pm12$ \\
\hline
\multicolumn{1}{r}{\hbox{CDF-S} Sources:} && $z_{850}<21$  & $z_{850}<22$  & $z_{850}<23$  & $z_{850}<24$  & $z_{850}<25$  & $z_{850}<26$  & all sources   & already   \\
\multicolumn{1}{r}{Number stacked:}&& $128$         & $289$         & $537$         & $1034$        & $1913$        & $3426$        & $9599$        & resolved  \\
\hline
& $0.5$--$1$                        & $1.9 \pm0.4 $ & $3.0 \pm0.6 $ & $4.5 \pm0.8 $ & $6.6 \pm1.1 $ & $9.0 \pm1.5 $ & $12.7\pm2.0 $ & $17.4\pm3.1 $ & $74\pm9 $ \\
& $1$--$2$                          & $1.4 \pm0.2 $ & $2.8 \pm0.3 $ & $4.3 \pm0.4 $ & $6.4 \pm0.5 $ & $9.2 \pm0.7 $ & $12.1\pm0.9 $ & $19.1\pm1.4 $ & $72\pm5 $ \\
& $2$--$4$                          & $1.1 \pm0.5 $ & $2.4 \pm0.7 $ & $3.4 \pm0.9 $ & $6.4 \pm1.3 $ & $9.7 \pm1.8 $ & $10.0\pm2.3 $ & $19.4\pm3.8 $ & $83\pm7 $ \\
& $4$--$6$                          & $1.0 \pm0.9 $ & $1.2 \pm1.3 $ & $2.7 \pm1.8 $ & $3.0 \pm2.5 $ & $6.5 \pm3.4 $ & $4.7 \pm4.4 $ & $10.5\pm7.4 $ & $82\pm11$ \\
& $6$--$8$                          & $5.4 \pm5.1 $ & $9.9 \pm7.6 $ & $5.3 \pm10.4$ & $<42.1$       & $6.8 \pm18.9$ & $12.6\pm25.3$ & $32.4\pm42.3$ & $66\pm11$ \\
\hline
\end{tabular}
\end{table*}

We repeated the stacking analyses using sub-samples of the GOODS sources,
selecting on the $z_{850}$-band, which samples the \hbox{$\sim8300$--$9500\a$}
wavelength range. The publicly-available GOODS catalogue is $z_{850}$-band
selected. The strongest detection of a soft (\hbox{$0.5-2\kev$}) X-ray signal
occurs for the \hbox{$z_{850}\lesssim25$} galaxies, whilst the highest
signal-to-noise of hard X-ray signals tend to occur for the brighter
sub-samples, although the detections are borderline except for a
\hbox{$4$--$6\kev$} signal at \hbox{$4.3\sigma$} for the \hbox{CDF-N} with
\hbox{$z_{850}<23$} sources. There is also hint of a \hbox{$6$--$8\kev$} signal
for the \hbox{CDF-N} with \hbox{$z_{850}<23$} sources (\hbox{$1.2\sigma$}), and
the \hbox{CDF-S} with \hbox{$z_{850}<22$} sources (\hbox{$1.9\sigma$}). For the
individual sources, we plotted X-ray counts against optical flux in various
bands but saw no obvious correlation, nor was there any obvious trend apparent
in similar plots of X-ray flux against various optical colours.

\subsection{Source contributions to the unresolved XRB}

Table~\ref{countssummary} shows the total flux due to the stacked sources
expressed as a percentage of the total XRB level (refer to
section~\ref{xrbmodel} for details of the XRB model assumed). Around
\hbox{$\sim15$--$20$} per cent of the \hbox{$0.5$--$4\kev$} background is
contributed by the GOODS sources. In the \hbox{$4$--$6\kev$} band the
contribution is \hbox{$\sim10\pm5$} per cent. However, only a \hbox{$3\sigma$}
upper limit of \hbox{$82$} per cent can be placed on the \hbox{$6$--$8\kev$}
contribution using the \hbox{CDF-N}, with the \hbox{CDF-S} indicating
\hbox{$32\pm42$} per cent. Both measurements are consistent with both zero and a
large contribution in this range.

\begin{figure*}
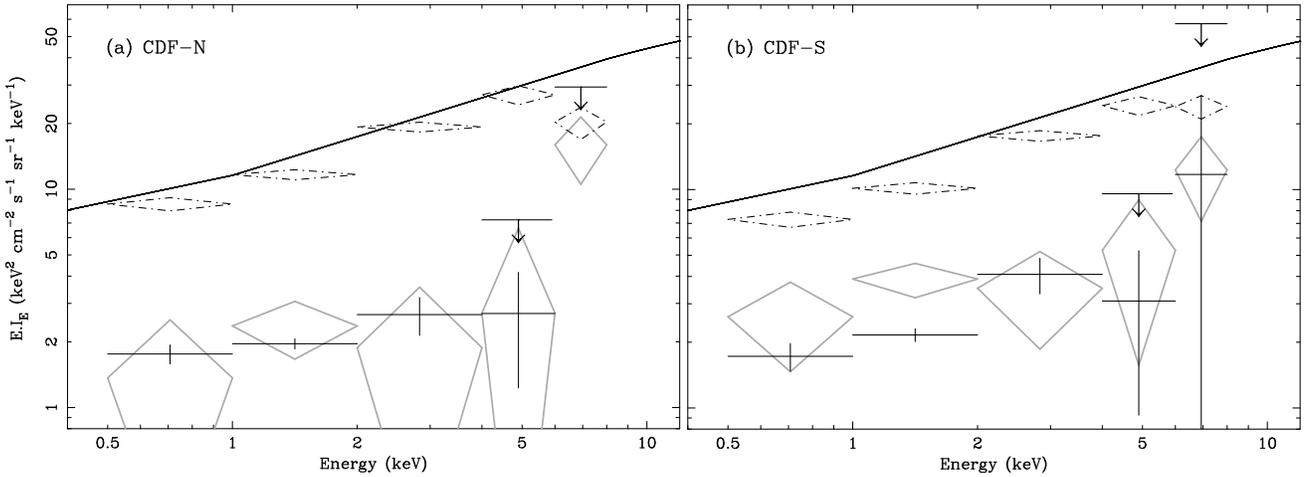

\rotatebox{270}{
\resizebox{!}{9cm}
{\includegraphics{nuinu_cdfn.ps}}
}
\rotatebox{270}{
\resizebox{!}{8.18cm}
{\includegraphics{nuinu_cdfs.ps}}
}
\caption{The extragalactic X-ray background with the resolved and stacked
  contributions. Panels (a) and (b) shows the data for the \hbox{CDF-N} and
  \hbox{CDF-S} respectively.  In each case the solid black line shows the total
  XRB level and the broken black diamonds show the intensity resolved from
  individually-detected X-ray sources (see Introduction and
  \citealp{worsley05}). The grey diamonds indicate the missing intensity, i.e.
  the residual remaining once the resolved contributions are subtracted from the
  total XRB level. The solid black crosses now indicate the stacked intensity
  due to the optically-detected, but individually X-ray undetected, GOODS
  sources determined in the work here. Additionally, for the \hbox{$4$--$6$} and
  \hbox{$6$--$8\kev$} bands, $3\sigma$ upper limits to the stacked intensity are
  shown.}
\label{nuinu}
\end{figure*}

Fig.~\ref{nuinu} shows the spectral shape of the stacked X-ray emission from the
GOODS sources, and the relative contributions to the total XRB level. It is
encouraging to note that (at least for the \hbox{$<6\kev$} range) we find the
stacked contribution to the X-ray background agrees well with the `missing'
intensity once that from X-ray detected sources has been considered. The large
errors mean that there remains some room for other emission, particularly in the
\hbox{$0.5$--$2\kev$} band, even when a \hbox{$\sim6$} per cent contribution
from clusters is taken into account \citep[see e.g.][]{moretti03}. The residual
emission can place important constraints on any truly diffuse components
\citep{soltan03} and the total contribution of accreting black holes to
reionization \citep{dijkstra04} but it is difficult to place any limits on the
still-unresolved emission given the large errors and without a more thorough
understanding of cosmic variance.

The reduction in sensitivity of \chandra\ above \hbox{$6\kev$} means that the
instrumental background heavily outweighs any source signal and errors in the
stacked X-ray intensity become very large. In the \hbox{CDF-N}
\hbox{$6$--$8\kev$} band the detection of a stacked signal representing
\hbox{$\sim40$} per cent of the total XRB -- the full unresolved intensity in
this band -- would only correspond to a \hbox{$\snr\sim2.1\sigma$}
(\hbox{$\snr\sim1.4\sigma$} in the \hbox{CDF-S}), with background counts
exceeding source counts by a factor \hbox{$\sim100$}. The \hbox{$3\sigma$} upper
limit of \hbox{$82$} per cent to the \hbox{CDF-N} stacked intensity is easily
consistent with what is required to account fully for the background. The
\hbox{CDF-S} stacked intensity gives \hbox{$32\pm42$} per cent although the
error is again very large. The \hbox{$3\sigma$} upper limit in the \hbox{CDF-S}
exceeds \hbox{$100$} per cent.

Since GOODS will detect \hbox{$L_{\star}$} galaxies out to redshift
\hbox{$z\sim6$} we would expect the optically-detected sources to include almost
every galaxy capable of hosting a significant AGN. Although the stacked X-ray
flux from the GOODS optical sources is consistent with the missing XRB
intensity, it is still possible that the contribution could be much lower and we
briefly discuss the possible source populations which may be missing from our
analysis. If the actual contribution is indeed much less than is needed to
explain the unresolved XRB, then the missing AGN must come from rare sources,
occurring with sky densities \hbox{$\lesssim10\pdegsq$} and so not sampled well
in the pencil-beam surveys; or, from very faint galaxies that are not optically
detected in the GOODS survey.

\subsubsection{Rare source populations}

The first of these possibilities requires the missing AGN to occur with sky
densities \hbox{$\lesssim10\pdegsq$}, therefore not sampled by the
\hbox{$\sim0.1\degsq$} deep surveys, and explaining why the stacked source
emission fails to account for \hbox{$\sim40$} per cent of \hbox{$6$--$8\kev$}
XRB emission. However, it is also clear that the missing sources must occur with
sky densities \hbox{$\gtrsim1\pdegsq$}. This is because total XRB emission
measurements by \citet{deluca04}, who analyse the background in the fields of
\hbox{$\sim5.5\degsq$} of \xmm\ observations, find the total XRB emission is a
\hbox{$\Gamma=1.4$} power-law up to at least \hbox{$8\kev$}. The missing AGN
must therefore be present at \hbox{$\gtrsim1\pdegsq$} in order to be present in
their analysis and provide the full XRB flux.

At densities of \hbox{$1$--$10\pdegsq$}, in order to be responsible for the full
\hbox{$\sim40$} per cent of unresolved \hbox{$6$--$8\kev$} XRB, the required
flux per source would need to be \hbox{$\gtrsim2\times10^{-13}\ergpcmsqps$} in
this band. Since only \hbox{$\sim10$} per cent of the \hbox{$4$--$6\kev$} band
is unresolved \citep{worsley05}, the required sources need to be Compton thick
with \hbox{$N_{\rm{H}}\gtrsim2\times10^{24}\pcmsq$} (to have a negligible
\hbox{$\lesssim6\kev$} contribution) and of quasar luminosity
(\hbox{$L_{2-10\kev}\gtrsim5\times10^{44}$}). Whilst \citet{gandhi04} have found
several type II QSOs serendipitously at sky densities of
\hbox{$\gtrsim10\pdegsq$}, their sources tend to have
\hbox{$N_{\rm{H}}=10^{23}$--$10^{24}\pcmsq$} and the contribution to the XRB in
the \hbox{$6$--$8\kev$} band is only few per cent.

\subsubsection{Optically undetected sources}

The second alternative -- that AGN hosts could remain optically undetected -- is
supported by the discovery of extreme X-ray/optical sources (EXOs) which show
X-ray fluxes of \hbox{$\sim10^{-16}$--$10^{-15}\ergpcmsqps$}, yet are optically
undetected with \hbox{$z_{850}\gtrsim28$} \citep{koekemoer04a}. The 7 EXOs in
the \hbox{CDF-S} have now been detected using \hbox{$24\um$} \spitzer\
observations \citep{koekemoer04b}. Interestingly, \citet{wang04} find that
several of the EXOs show very hard X-ray spectra, consistent with type II AGN at
redshift \hbox{$\lesssim6$} hosted by very under-luminous, or very dusty
galaxies. It is plausible that X-ray undetected EXOs may contribute a few per
cent to the total XRB level (those identified to date in the \hbox{CDF-S}
contribute \hbox{$\sim0.5$--$1.5$} per cent).

\subsection{Spectral shape: AGN and starburst components}

The softer \hbox{$\lesssim4\kev$} flux seen from the GOODS sources is most
likely to be dominated by starburst emission in these faint galaxies with a
photon index \hbox{$\Gamma\sim1.5$--$2$}. The average \hbox{$0.5$--$2\kev$} flux
of the GOODS sources is \hbox{$\sim2\times10^{-18}\ergpcmsqps$} at a sky density
of \hbox{$\sim6\times10^{5}\pdegsq$}. Adding this point to a soft X-ray
\hbox{$\log N$--$\log S$} diagram \citep[see fig. 4 in][]{bauer04} is consistent
with a direct extrapolation of the steeply-climbing number-counts distribution
of star-forming galaxies, which at these very faint fluxes, outnumber AGN by an
order of magnitude. Starburst activity is a likely explanation for the
\hbox{$\lesssim4\kev$} stacked emission although weak AGN activity is also
possible.

\citet{worsley05} model the missing hard XRB intensity as emission from a
population of heavily obscured AGN: a simplified spectral model consisting of a
\hbox{$\Gamma=2$} power-law, plus photoelectric absorption and a reflection
component, is able to explain the shape of the missing \hbox{$>2\kev$} XRB
emission inferred from the \chandra\ Deep Fields for redshifts
\hbox{$\sim0.5$--$1$} and column densities
\hbox{$\sim10^{23}$\thinspace--\thinspace$5\times10^{24}\pcmsq$} (see
Fig.~\ref{countssummary}). Within the errors, the stacked X-ray emission from
the GOODS sources does have a spectral shape which is consistent with that
predicted from obscured AGN activity.

Assuming that the missing hard XRB can be explained by the hypothesised
population of obscured AGN, the difficulty is in reducing the error in the
\hbox{$6$--$8\kev$} stacked \chandra\ emission to the point where definite
verification or inconsistency can be seen between the XRB residual and the
stacked GOODS intensity. If, for example, only $10$ per cent of the GOODS
galaxies contain an obscured AGN, then the overwhelming majority of stacked
X-ray positions are simply adding background noise to the measurement; we need
to be able to select the likely AGN candidates. Our attempt to do this on the
basis of the $z_{850}$-band optical magnitude was not successful, with no
particular $z_{850}$-band magnitude restriction improving the
\hbox{$6$--$8\kev$} signal over that obtained when all sources were included,
although there is a weak improvement when only the brighter
(\hbox{$z_{850}\lesssim24$}) objects are selected. Selections of
\hbox{$z_{850}\lesssim23$--$25$} did, however, succeed in increasing the \snr\
of the \hbox{$4$--$6\kev$} band considerably: from \hbox{$2.6\sigma$} to
\hbox{$4.3\sigma$} in the \hbox{CDF-N} ($z_{850}\lesssim23$); and from from
\hbox{$2.0\sigma$} to \hbox{$2.8\sigma$} in the \hbox{CDF-N}
($z_{850}\lesssim25$).

A much better method of discrimination could be the use of infrared fluxes which
will soon be available from \spitzer\ observations. Highly absorbed AGN are
expected to show strong far infrared emission due to the reprocessing of X-ray
emission by the obscuring dust. The difficulty here would be separating the
obscured AGN candidates from strong starburst galaxies, which are also bright in
the infrared.

\section{Conclusions}

We have stacked the emission from X-ray undetected optical sources in the GOODS
fields to address the question of whether they can account for the missing hard
XRB implied by the stacking of X-ray detected sources \citep{worsley05}. Our
central conclusion is `plausibly, but not certainly, yes'. Whilst X-ray emission
is detected at high significance in the \hbox{$0.5$--$4\kev$} range, the strong
downturn in the \chandra\ sensitivity makes detection more difficult at higher
energies. By selecting the sources on the basis of $z_{850}$-band
(\hbox{$8300$--$9500\a$}) magnitude, we were able to obtain \hbox{$4.3\sigma$}
and \hbox{$2.8\sigma$} detections of \hbox{$4$--$6\kev$} emission in the
\hbox{CDF-N} and \hbox{CDF-S} fields respectively, whilst in the
\hbox{$6$--$8\kev$} band we only achieve constraints at the \hbox{$1.2\sigma$}
and \hbox{$1.9\sigma$} levels, respectively.

The \hbox{$0.5$--$4\kev$} emission represents some \hbox{$15$--$20$} per cent of
the XRB with \hbox{$\sim10\pm5$} per cent in the \hbox{$4$--$6\kev$} band. When
added to the fractions due to the resolved XRB the totals are consistent with
\hbox{$100$} per cent, although depending on XRB normalisation (and spectral
shape \hbox{$<1\kev$}), there remains room for contributions from other sources.
The \hbox{CDF-S} \hbox{$6$--$8\kev$} intensity is \hbox{$32\pm42$} per cent,
suggesting that at least half of the unresolved XRB (\hbox{$\sim40$} per cent),
may be lurking in these sources; however, the errors are considerable, and the
\hbox{CDF-N} shows no clear excess in the stacked \hbox{$6$--$8\kev$} intensity
above the instrumental background, although the \hbox{$3\sigma$} upper limit to
the potential XRB contribution is \hbox{$82$} per cent in the \hbox{CDF-N},
which is more than enough to account for the unresolved intensity.

In previous work, we predicted that the missing component of the XRB is due to
moderate luminosity, highly obscured AGN, with absorption column densities of
\hbox{$\sim10^{23}$\thinspace--\thinspace$5\times10^{24}\pcmsq$}, at redshift
\hbox{$\sim0.5$--$1.5$}. Since GOODS is able to detect \hbox{$L_{\star}$}
galaxies out to a redshift \hbox{$z\sim6$}, we expected it to include the hosts
of this heavily obscured AGN population. The stacked X-ray flux in the
\hbox{CDF-S} shows a larger contribution in the \hbox{$6$--$8\kev$} band (see
Fig.~\ref{nuinu}), precisely where the steeply-rising spectrum of these sources
becomes important. This could be the sought-after signature of the missing
population, although the poor \snr\ and the consequently large errors mean the
detection is uncertain.

Our results are also consistent with a negligible additional contribution to the
XRB in the \hbox{$6$--$8\kev$} band, which if true, would imply that the highly
obscured AGN population is hosted by galaxies which are optically-undetected in
the GOODS fields. Even with the recent discovery of EXOs, it seems unlikely that
such a large AGN population can be accommodated by such optically faint hosts. A
strong contribution could also be due to Compton thick quasars occurring with sky
densities of \hbox{$1$--$10\pdegsq$}.

In order to boost the signal-to-noise of the stacked GOODS signal there remains
much work to be done in selecting the appropriate sub-sample of sources, and
avoid adding background noise to the measurement by including ordinary galaxies.
Infrared-bright galaxies identified by \spitzer\ may hold the key to doing this,
assuming differentiation between dusty star-forming galaxies, and potential
obscured AGN, can be accomplished. Additional deep field observations,
particularly further exposure in the \hbox{CDF-N}, could resolve the situation
by increasing the sensitivity to individual, faint X-ray sources, as well as
improve in the signal-to-noise ratio in stacked-source signals in the hardest
energy bands. Additional investigations into EXOs will also provide useful
constraints, as will shallow wide-area surveys at hard X-ray energies, which can
place important limits on the contribution from heavily obscured quasars.

As a final comment we note that there is good agreement with the recent work of
\citet{hickox05} where both the resolved sources and unresolved background are
measured in the \hbox{$0.5$--$8\kev$} band using \chandra\ in the Deep Fields.
The total background they determine has a normalization 6 per cent less than we
assume (Section~\ref{xrbmodel}), but with an uncertainty of 11 per cent. A large
amount of the difference in normalisations is probably due to differences in
analysis methods, the most significant of which is our `bright end correction'
\citep[see][]{worsley05} which attempts to correct for the poor sampling of the
bright end of the X-ray number counts distribution. The resolved fractions found
by Hickox \& Markevitch are in good agreement with our results in the bands used
(\hbox{$1$--$2$} and \hbox{$2$--$8\kev$}).

\section{Acknowledgments}

MAW and FEB acknowledge support from PPARC. ACF and DMA thank the Royal Society
for support. WNB and BDL thank NSF career award AST-9983783 and CXC grant
GO4-5157 for support. MAW would like to thank his coauthors for their support
and patience in seeing this work through to publication.

\bibliographystyle{mnras} 
\bibliography{mn-jour,worsley_26feb06}

\end{document}